\documentclass[reprint,aps,prx,twocolumn,showpacs,superscriptaddress,10pt]{revtex4-1}

\usepackage{amsmath}
\usepackage{amsfonts}
\usepackage{amssymb}
\usepackage{graphicx}
\usepackage{color}
\usepackage{soul}
\usepackage{xcolor}
\usepackage[colorlinks=true,linkcolor=blue]{hyperref}
\usepackage{braket}

\graphicspath{{./figures/}}

\begin{document}
\title{
Simulating a quantum commensurate-incommensurate phase transition \\using two Raman coupled one dimensional condensates}

\author{V. Kasper}
\affiliation{Department of Physics, Harvard University, Cambridge, MA, 02138, USA}
\author{J. Marino}
\affiliation{Department of Physics, Harvard University, Cambridge, MA, 02138, USA}
\affiliation{Institut f\"ur Physik, Johannes Gutenberg Universit\"at Mainz, D-55099 Mainz, Germany}
\author{S. Ji}
\affiliation{Vienna Center for Quantum Science and Technology, Atominstitut, TU Wien, Stadionallee 2, 1020 Vienna, Austria}
\author{V. Gritsev}
\affiliation{Institute for Theoretical Physics, Universiteit van Amsterdam, Science Park 904, Postbus 94485, 1098 XH Amsterdam, The Netherlands}
\affiliation{Russian Quantum Center, Skolkovo, 121205 Moscow region, Russia}
\author{J. Schmiedmayer}
\affiliation{Vienna Center for Quantum Science and Technology, Atominstitut, TU Wien, Stadionallee 2, 1020 Vienna, Austria}
\author{E. Demler}
\affiliation{Department of Physics, Harvard University, Cambridge, MA, 02138, USA}

\begin{abstract}
We study a transition between a homogeneous and an inhomogeneous phase
in a system of one-dimensional, Raman tunnel-coupled Bose gases. The
homogeneous phase shows a flat density and phase profile, whereas the
inhomogeneous ground state is characterized by a periodic density
ripples, and a soliton staircase in the phase difference. We show that under experimentally viable conditions the transition can be tuned by the wavevector difference $Q$ of the Raman beams and can be described by the Pokrovsky-Talapov model for the relative phase between the two condensates. Local imaging available in atom chip experiments allows to observe the soliton lattice directly, while modulation spectroscopy can be used  to explore collective modes, such as the phonon mode arising from breaking of translation symmetry by the soliton lattice.  In addition, we investigate  regimes where the cold
atom experiment  deviates from the Pokrovsky-Talapov
field theory. 
We predict unusual mesoscopic effects arising from the finite size of the system, such as quantized injection of solitons upon increasing $Q$, or the system size. 
For moderate values of $Q$ above
criticality, we find that the density modulations in the
two gases interplay with the relative phase profile
and introduce novel features in the spatial structure
of the mode wave-functions. Using an inhomogeneous
Bogoliubov theory, we show that spatial quantum
fluctuations are intertwined with the emerging
soliton staircase.  Finally, we comment on the
prospects of the ultra-cold atom setup as a
tunable platform studying quantum aspects of
the Pokrovsky-Talapov theory in and out-of-equilibrium.
\end{abstract}

\date{\today}
\maketitle

\section{Introduction}
\begin{figure}[t!]
\includegraphics[scale=.292]{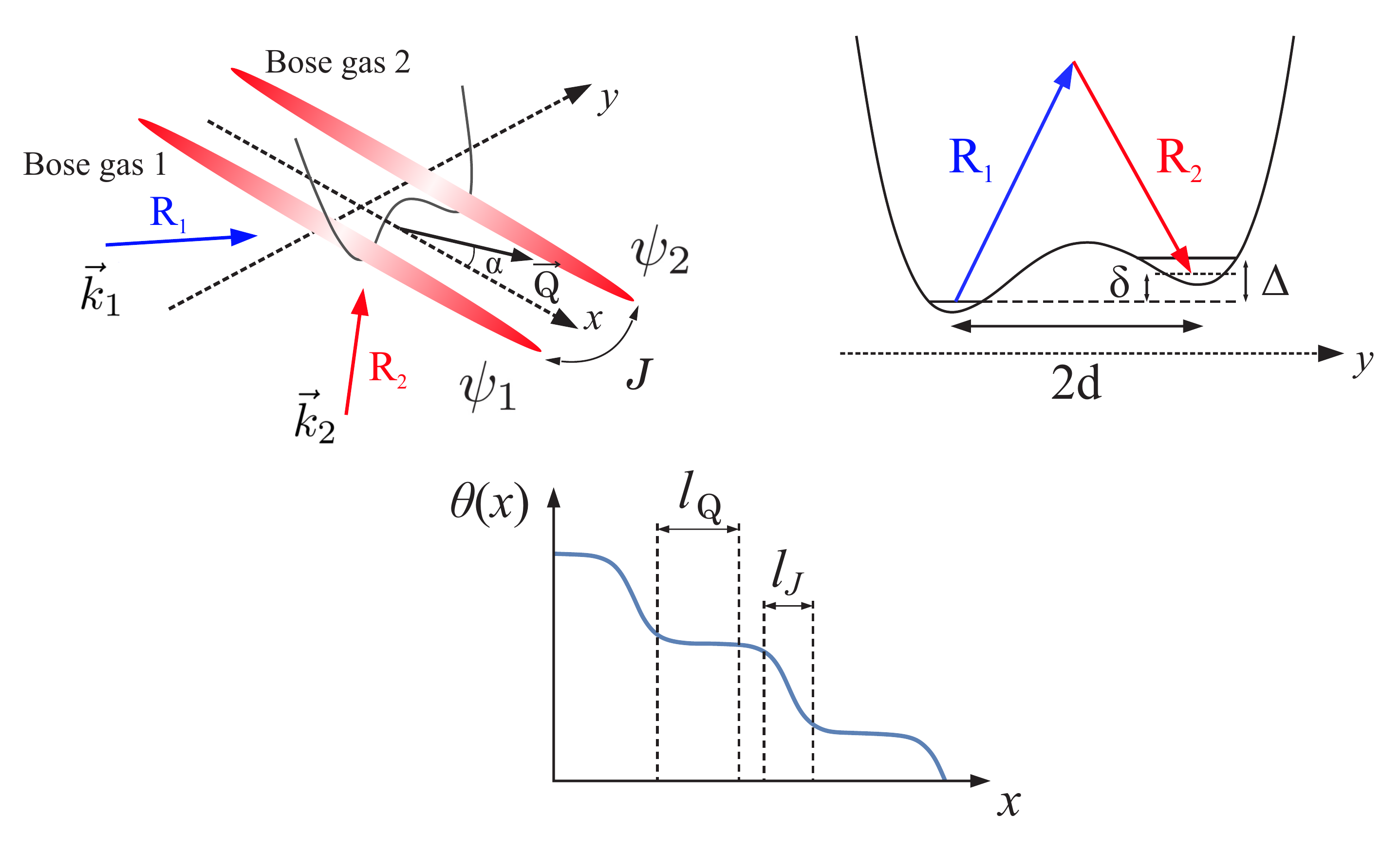}
\caption{\textbf{Experimental platform: Raman tunnel coupled Bose gases} Two Raman laser beams R1 and R2 with wave-vectors $\vec{k}_1$ and $\vec{k}_2$ induce a  spatially-modulated tunnel coupling between two Bose gases $\psi_1(x)=\sqrt{n_1(x)}e^{i\theta_1(x)}$ and $\psi_2(x)=\sqrt{n_2(x)}e^{i\theta_2(x)}$. The detuning $\delta$ prevents direct resonant tunneling between the two wells. The angle $\alpha$ between the two Raman beams allows to tune the mismatch wave vector $\vec{Q}=\vec{k}_1-\vec{k}_2$. The phase difference of the two condensates $\theta(x)= \theta_1(x)-\theta_2(x)$ has the shape of a staircase upon increasing $Q$ above a critical threshold. The length scale $l_J$ on which the phase jump of $2\pi$  takes place, is determined by the tunnel coupling $J$ and proportional to $1/\sqrt{J}$. By changing $Q$ one can modify the length  between two adjacent solitons  $l_Q\propto 1/Q$, see ~\cite{arist} and~\cite{laz}.  \label{figplat} }
\end{figure}
Ultracold atom and molecule systems serve as versatile platforms for precision measurements \cite{Oelker2019}, quantum computation \cite{DeMille2002} and quantum simulation \cite{Mazurenko2017}. In particular ultracold atom systems have several features which make them particularly well suited for studying many-body physics. Their isolation from the environment is excellent and allows the observation of coherent quantum evolution undisturbed by coupling to external baths \cite{Bernien2017}. In systems of ultracold atoms microscopic parameters are not only highly tunable but they can also be changed on time-scales that are much shorter than typical time-scales of many-body systems. The rich toolbox of atom physics, as for example imaging with single-atom resolution, enables a detailed characterization of quantum many-body states which is difficult to achieve in any other system \cite{Lukin2019}. 
Current research is highly enriched by the unique perspective ultracold atoms provide for the study of many-body problems. On the one hand they can be used to realize paradigmatic models of quantum many-body systems/quantum simulators such as Bose- and Fermi-Hubbard models \cite{Greiner2002}, band structures with geometrical and topological features \cite{Stuhl2015}, one dimensional Luttinger liquids \cite{Gring2012} or the sine-Gordon model \cite{Jorg15}. Ultracold atoms provide us also with examples of quantum systems that do not have analogues in high energy or solid state physics. For example, spinor bosonic atoms realise a wide variety of spinor condensates~\cite{imambekov2003spin} and Mott states~\cite{duan2003controlling} with magnetic order~\cite{Kawaguchi2012}. Alkaline-earth atoms provide realizations of Fermi Hubbard models with SU($N$) symmetry \cite{Gorshkov2010} with $N$ up to 10. 
Especially, the possibility to study phase transition with quantum simulators has the prospect to settle fundamental questions such as the binding mechanism of Cooper pairs in the Fermi-Hubbard model or to study strongly coupled quantum critical points. 

Another important, yet less explored, class of phase transitions is given by commensurate-incommensurate phase transitions. A famous example  is given by the adsorption of atoms on a crystalline substrate e.g. rare-gas monolayers adsorbed on graphite~\cite{Bak}. This system can be described by a discrete version of the Frenkel-Kontorowa model which is a one-dimensional chain of atoms connected by springs and are subject to a cosine potential. If the cosine potential dominates, the adsorbed atoms relax into a commensurate structure where the average lattice spacing of the adsorbed atoms is a rational fraction of the period of the periodic potential, while for a weaker potential the adsorbed atoms form an incommensurate structure.
Another example for commensurate-incommensurate phase-transitions arises in conductors due to the interactions between conduction electrons and the atomic lattice. The conduction electron density is spatially modulated forming a charge density wave which in turn follows the periodic lattice distortion. Commensurate-incommensurate transitions also appear in magnetic systems like the anisotropic Ising model with competing interactions, type-II superconductors, quantum Hall bilayer systems under a tilted magnetic field~\cite{dro, schu, yan, moon, yan2}, and the transition from zero to finite momentum pairing in  two-component attractive fermions in presence of a Zeeman field~\cite{ov,book,RevModPhys.85.1633}. A comprehensive list of experimental platforms can be found in~\cite{Bak,pokrovsky1983theory,brazovskii2009solitons}. 

In this work we propose the realization of a commensurate-incommensurate phase transitions in a quantum platform of ultra-cold atoms confined to  one-dimensions, which in terms of tunability and flexibility excels traditional solid-state setups employed to realize commensurate structures. Specifically we consider a pair of tunnel-coupled, one-dimensional tubes of ultracold $^{87}$Rb atoms prepared in an elongated micro-trap on an atom-chip~\cite{schumm}, which has been recently shown to constitute an efficient  simulator of the quantum sine-Gordon model by measuring higher order correlation functions~\cite{Jorg15}. We revive the tunneling amplitude between the two one-dimensional Bose gases with Raman beams and modulate it spatially (see Fig.~\ref{figplat}). The spatial modulation corresponds to a situation in which upon tunnelling  atoms acquire a finite momentum  along the direction of the tubes.
This modification enlarges the potential of one-dimensional Bose gases on atom-chips as quantum simulators: imprinting a phase winding on the tunneling process constitutes a  realization of the Pokrovsky-Talapov (PT) model~\cite{ners, japaridze1979one, burkov, schulz, arist,  PT, PT2, Bak, laz, Giam, giamarchi1988theory,fendley2004competing}. This is a variant of the sine-Gordon field theory, which hosts a commensurate-incommensurate transition between a homogeneous and inhomogeneous phase. The latter supports the onset of solitons in the relative phase profile of the two gases and provides an avenue to engineer and manipulate topological structures in tunnel-coupled one-dimensional Bose gases on atom-chips. 
Specifically, the PT Hamiltonian is given by
\begin{equation}\label{hamintro}
H = \int dx \left[\frac{1}{2} (\partial_x \Theta)^2   -2 \cos \Theta   + Q \partial_x \Theta\right]  ,
\end{equation}
where $Q$ acts as a chemical potential for the phase gradient which in the inhomogeneous phase ($Q>Q_c=4\sqrt{2}/\pi$) imprints solitons in the ground state of the system~\cite{giamarchi2003quantum}. 

We will show that the field $\Theta$ in equation \eqref{hamintro} can be identified with the phase difference of the two one-dimensional Bose gases (cf. Fig.~\ref{figplat}). We shall see that the microscopic model describing our platform,  has several differences with the basic PT model in Eq.~\eqref{hamintro}, where the most important one is the coupling between the relative phase and the symmetric density of the two tubes. 
Such coupling between phase and density appears not important in other realizations of the commensurate-incommensurate transition in cold atoms simulators such as those based on optical lattices~\cite{zwerger,haller2010pinning}.

Further commensurability effects for cold fermionic atoms and strongly interacting bosons trapped in one dimensional optical lattices have  been studied in Refs.~\cite{ov, lazarides2011strongly,PhysRevB.77.245105}, while  commensurate-incommensurate phase transitions involving chains of cold trapped ions have been investigated in Refs.~\cite{garcia2007frenkel, fogarty2015nanofriction,bylinskii2016observation}.

\subsection{Summary of results}

We demonstrate that two one dimensional condensates with Raman assisted tunnelling of atoms provide a new experimental platform for studying the quantum commensurate-incommensurate transition. Local imaging available in atom chip experiments can be used to study key features of the model including formation of the soliton lattice in the incommensurate phase and appearance of the gapless phonon mode arising from breaking of translational symmetry by the soliton lattice. We also discuss differences between the microscopic Hamiltonian and the canonical PT model which become prominent    
 close to the critical region of the commensurate/incommensurate transition, or for small systems. In particular, we observe density ripples in the density profiles of the two gases which are in a one-to-one correspondence with kinks in their phase difference.  The onset of (i)~\emph{solitons in the density of the Raman tunnel coupled quantum liquids} is a result which goes beyond the description with the PT theory (see Sec.~\ref{sec:first}). These spatial modulations in the density profiles of the two gases is at the root of (ii)~\emph{discrepancies from the PT effective field theory description}.  Specifically, we compare the density of solitons in our simulator with density of solitons in the PT theory in  Fig.~\ref{fig:PD2} of Sec.~\ref{sec:EFT}. We find the largest deviations for small system sizes and misfit parameter close to criticality. The finite size study of our model illustrates novel   (iii)~\emph{mesoscopic effects} such as the  \emph{quantised injection of solitons} into the system, which are absent in the PT theory for an infinite system (see Fig.~\ref{fig:Injection}). Our cold atom implementation allows to study the spatial features of  quantum many-body states and comes with a level of tunability which is difficult to achieve in solid state platforms realising commensurate/incommensurate transitions, as for example via changing the lattice spacing constant. Some features of the commensurate/incommensurate phase transition can be understood with a classical treatment. However, the quantum nature of ultracold atom platform allows for the exploration of quantum effects in the incommensurate soliton structures. As an example we show in Sec.~\ref{sec:bogo} that (iv)~\emph{ the wave-functions of the excited states have peaks exactly at the position of solitons}. Therefore, the corrugated spatial profile of the modes is informative of the soliton structures shaped by the momentum imprinted by the Raman beam on the atoms.
We arrive at these results by using a variational approach and an inhomogeneous Bogoliubov theory. The latter is a straightforward approach   to include quantum fluctuations when compared to other  methods (Bethe ansatz ~\cite{hal, caux, papa, papa2}, semi-classical wavefunctions~\cite{arist}) which require to regularise  the PT field theory in the ultraviolet. 

\section{Experimental implementation  }
 \label{sec:MF}
In this section we discuss the setup of two one-dimensional Raman tunnel coupled Bose gases.
In atom chip experiments one can manufacture one-dimensional Bose gases by creating a magnetic trap, 
which is generated by electric currents on the atom-chip itself~\cite{Folman02, Reichel_Book11}.
In order to create a double well potential one uses a pair of radio-frequency  wires on the atom chip \cite{Nature2006Hofferberth, PRA2006Lesanovsky}:
the radio frequency field will dress the magnetic states of the atoms and time-averaging results in an effective double well potential. By tuning the strength of the radio-frequency fields one can vary the distance between the double wells changing in this way the Josephson tunneling strength $J_0$.

We also introduce a tilted   double well potential which leads to an energy offset $\Delta$ and suppresses the direct tunneling process.
In an atom chip experiment this tilting can be achieved by changing the ratio of the different radio frequency currents.
The coupling between the left and right well can be revived using Raman assisted tunneling~\cite{Jak_2003, Aid_2011,Miy_2013}. One adds two laser beams,
which create the potential 
$V_{R}(\vec{r})=V_0(r)\cos(\delta \omega \cdot t + \vec{Q}\cdot \vec{r})$
with a slight detuning $\delta \omega$ and the mismatch of the wave vectors  $\vec{Q} = \vec{k}_1 - \vec{k}_2$ of the two laser beams.
The effect of the Raman beams on the atoms in the
left and right tube can be modeled by the following Hamiltonian matrix
\begin{equation}
H=
{\left( \begin {array} {cc}
V_+(x,t) & \hbar J_0\\
\hbar  J_0 & V_-(x,t)-\hbar \Delta
\end {array}
\right )},
\end{equation}
with
\begin{align}
    V_{\pm}(x,t) = V_{0}\sin(\delta \omega t\pm Q_y d+ Q_x x),
\end{align}
where the distance between the double well is given by $2d$. 

The detuning of the Raman beams is chosen such that it 
equals the energy difference between the double well i.e. $\delta \omega \simeq \Delta$. Going to the interaction picture with the unitary transformation
\begin{equation}
U=\exp
{\left( \begin {array} {cc}
-\frac{i}{\hbar\delta \omega}V_+(x,t)  & 0\\
0 & -\frac{i}{\hbar\delta \omega}V_-(x,t)-i\Delta t
\end {array}
\right )}
\end{equation}
and a subsequent time averaging leads to the effective Hamiltonian
\begin{equation}
H=
{\left( \begin {array} {cc}
0 & Je^{iQ_x x}\\
Je^{-iQ_x x} & 0
\end {array}
\right )} \, .
\end{equation}
In the last equation we have introduced the effective tunneling
\begin{equation}
\begin{aligned}
J=- J_0 \cdot \mathcal{J}_1 \left( \frac{2V_{0}\sin(Q_y
d)}{\hbar\Delta}\right) ,
\end{aligned}
\end{equation}
which is renormalized by a  Bessel function  of the first kind $ \mathcal{J}_1(\cdot)$ with index $n$. 

\subsection{Parameters for an atom chip experiment}
In this subsection we elaborate on the experimental 
details  and 
the possibilities to tune  parameters of a typical atom chip experiment.  
Starting from the confinement of the clouds 
the radial and longitudinal trapping frequencies are $\omega_\bot\approx2\pi\times1.4$kHz and $\omega_x\approx2\pi\times7$Hz, respectively. The distance between the wells 
can be continuously varied in the range from $\sim\!2$ to $3\,\mu\text{m}$, which allows to adjust the tunneling strength $J_0/(2\pi \hbar)$ from 0 to 100 Hz.
In addition to the magnetic trapping one can create a box potential by using a blue detuned laser beam, which cuts off the left and the right tails of the Bose gas. This blue 
detuned laser beam enables to change the length of the system between 30 and $60\,\mu\text{m}$ \cite{Science2018Rauer}. 
Digital mirror devices permit to create an additional dipole potential and achieve nearly arbitrary potential configurations. In this way we can generate a flat bottom box potential by compensating the longitudinal harmonic effect and even extend the length of the box up to $250\,\mu \text{m}$~\cite{Tajik19}.

For two Raman lasers with $\lambda=1064\,\text{nm}$ one
can achieve $\sin(\Delta Q_yd)\sim1$.
An intensity of $100\, \text{mW}/\text{mm}\vphantom{m}^2$ generates a dipole potential with a depth of $2\pi \hbar \times 100\,\text{Hz}$, which enables to change the amplitude $V_0/(\hbar  \Delta)$ from 0 to 0.5. This allows for an effective tunneling strength $|J|/J_0$ between $0$ and $0.4$ and the misfit parameter $Q_x$ can be varied between $0$ and $11.8\,\mu m^{-1}$ by changing the angle $\alpha$ between Raman beams, see Fig.\ref{figplat}.
Since the resolution of the imaging system is $\sim2\, \mu \text{m} $, we will only consider a maximum $Q_x$ of $0.628 \, \mu \text{m}^{-1}$ where we expect a $2\pi$ phase flip in the range of 5 pixels. In order to avoid unwanted radial excitations the Raman beams will be ramped slowly on a time scale of $10$~ms.

The atom density can be set between 
$n=10-100\mu \text{m}^{-1}$ which results in a chemical potential of $\mu=2\pi\times (0.05-0.5)\hbar \omega_\bot$
respectively. For a typical atom chip experiment with 
$^{87}$Rb the gas can be cooled down to below
  $ 10 \text{nK}$, which gives a thermal coherence length of $\lambda_T \sim 60\mu\text{m}$.
  This particle number and temperature allow to achieve an effective one-dimensional system because of $\mu, k_BT<\hbar \omega_\bot$. 
For a typical Josephson tunnelling $J/(2\pi\hbar)=1 \text{Hz}$ , the  Josephson length is $\lambda_J=(\hbar/4m{J})^{1/2}\approx
5\mu$m which is below the thermal coherence length $\lambda_J \ll \lambda_T$; this will allow to neglect thermal fluctuations in the following.
\section{Microscopic model}

We model the physical system presented in Fig.~\ref{figplat} with the Hamiltonian
\begin{align}
H =& \sum_{b=1}^2 \int_{0}^{L} dx \left[ \frac{\hbar^2}{2M} |\nabla\psi_b|^2 - \mu \psi^{\dagger}_b \psi_b   +  \frac{g}{2} \psi^{\dagger}_b \psi^{\dagger}_b \psi_b \psi_b \right] \notag \\
 &- \hbar J \int_{0}^{L} dx \left[ e^{iQx}\psi^{\dagger}_1 \psi_2 + e^{-iQx}\psi^{\dagger}_2 \psi_1 \right],  \label{eq:HMicro}
\end{align}
with $\mu$ the chemical potential of the two gases,  $J$ the tunneling amplitude, $g$
the interaction strength and $Q$ the mismatch wave vector.
The bosonic field operators  $\psi_{1}$ and $\psi_{2}$ fulfill
the  boundary conditions
\begin{align}\label{eq:BC}
\psi_b(x=0) = \psi_b(x=L) = 0 \,
\end{align}
due to the box potential.
In particular, the tunneling term $(e^{iQx}\psi^{\dagger}_1 \psi_2 + h.c)$ describes the tunneling of an atom between the two tubes while picking up a phase $\pm iQx$. In the following sections, we  choose typical values of $g$, $\mu$, $J$ and $Q$ relevant for experimental realizations with atom chip experiments ~\cite{schumm, Gring2012, Jorg15}.

\subsection{Ground state and phase transition}
\label{sec:first}
In this subsection we determine the ground state of the Hamiltonian given in \eqref{eq:HMicro} as 
a function of the
mismatch parameter $Q$.
 We employ the variational principle and choose the following coherent state ansatz 
\begin{align}
\ket{\alpha} = \prod_x e^{\alpha_1(x) \psi_1^{\dagger}(x)} e^{\alpha_2(x) \psi_2^{\dagger}(x)} \ket{0}
\end{align}
with the coherent state amplitudes $\alpha_1$ and $\alpha_2$.
The variational principle minimizes the expectation value of the energy
\begin{align}
E =  \frac{ \braket{\alpha|H|\alpha}}{\braket{\alpha|\alpha}},
\end{align}
which is given by
\begin{align}
E[\alpha] &= \sum_{b=1}^2 \int_{0}^{L} dx \left[ \frac{\hbar^2}{2M}|\partial_x\alpha_b|^2
- \mu |\alpha_b|^2   +  \frac{g}{2} |\alpha_b|^4  \right] \notag \\
 &- \hbar J \int_{0}^{L} dx \left[ e^{iQx}\alpha^{\ast}_1 \alpha_2 + e^{-iQx}\alpha^{\ast}_2 \alpha_1 \right] \, . \label{eq:Energy1}
\end{align}
The boundary conditions of the field operators in eq.~\eqref{eq:BC} imply that the coherent state
amplitudes  vanish at the edges of the system
\begin{align}
\alpha_b(x=0) = \alpha_b(x=L) = 0.  \label{eq:BCMicro}
\end{align}
In order to work with dimensionless quantities we rescale  space and field variables
\begin{align}
x &= x_0 \bar{x} \, , \\
\alpha_b &= x^{-1/2}_0 \bar{\alpha}_b \, ,
\end{align}
with the Josephson length
\begin{align}
x_0 = \sqrt{\frac{\hbar}{2MJ}}.
\end{align}
Given this rescaling the energy has the form
\begin{align}\label{en}
E[\bar{\alpha}] &= \sum_{b=1}^2 \int_{0}^{\bar{L}} d{\bar{x}} \left[ \frac{\hbar^2x^{-2}_0}{2M} |\partial_{\bar{x}}\bar{\alpha}_b|^2
- \mu |\bar{\alpha}_b|^2   +  \frac{g}{2x_0} |\bar{\alpha}_b|^4  \right] \notag \\
 &- \hbar J \int_{0}^{\bar{L}} d{\bar{x}} \left[ e^{i\bar{Q} u} \bar{\alpha}^{\ast}_1 \bar{\alpha}_2 + e^{-i \bar{Q} \bar{x}}\bar{\alpha}^{\ast}_2 \bar{\alpha}_1 \right] \, ,
\end{align}
where we have introduced the dimensionless length $\bar{L} = L/x_0 $ and wave vector
$\bar{Q} = x_0Q$. We measure the energy in Eq. \eqref{en} in units of $E_0 = {\hbar^2x^{-2}_0}/({2M})$
and hence the classical Hamiltonian reads
\begin{align}\label{eq:lattice}
\bar{E}[\bar{\alpha}] &= \sum_{b=1}^2 \int_{0}^{\bar{L}} d{\bar{x}} \left[  |\partial_{\bar{x}}\bar{\alpha}_b|^2
- \bar{\mu} |\bar{\alpha}_b|^2   +  \frac{\bar{g}}{2} |\bar{\alpha}_b|^4  \right] \notag \\
 &-  \int_{0}^{\bar{L}} d\bar{x} \left[ e^{i\bar{Q} \bar{x}} \bar{\alpha}^{\ast}_1 \bar{\alpha}_2 + e^{-i \bar{Q} \bar{x}}\bar{\alpha}^{\ast}_2 \bar{\alpha}_1 \right] \,,
\end{align}
where we introduced
$\bar{E}[\bar{\alpha}] =   {E[\bar{\alpha}] } / {E_0} $, the dimensionless chemical potential $\bar{\mu} = {\mu}/{E_0} $ and the dimensionless interaction strength $\bar{g}  = {g}/{(E_0 x_0)} $. From now on we will only work with dimensionless
quantities and drop the bar from the rescaled parameters unless differently stated.

\begin{figure*}
\includegraphics[scale=.45]{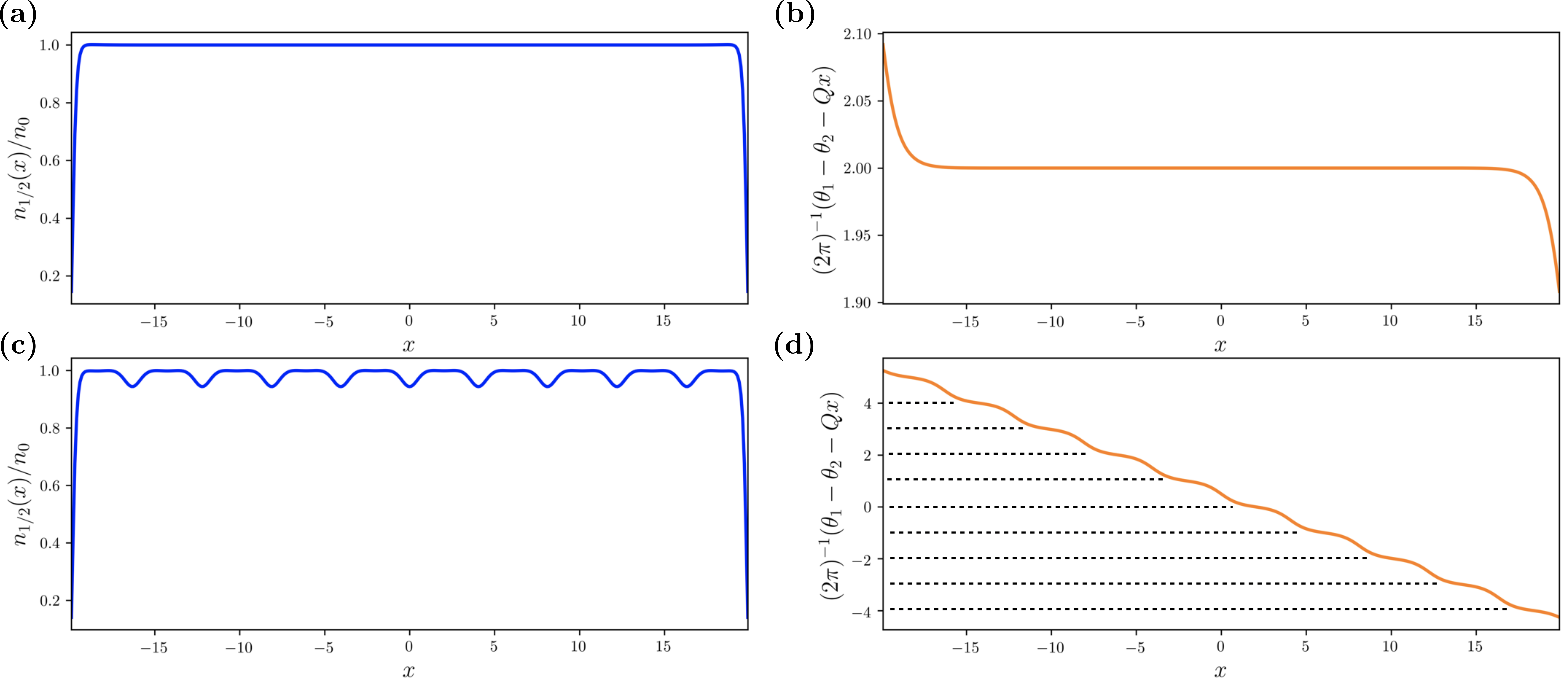}
\caption{\textbf{Density and shifted relative phase  profiles in the commensurate and incommensurate phase}: For a fixed system size of $\bar{L}=40$ and $\bar{Q}=0.7$ the density profile \textbf{(a)} as well as the relative phase profile \textbf{(b)} are homogeneous.  The microscopic parameters  are $\bar{\mu} = 30$ and $\bar{g} = 0.6 $ and are typical for atom chip experiments~\cite{schumm, Gring2012, Jorg15}. For a fixed system size of $\bar{L}=40$ and $\bar{Q}=1.9$ the density profile \textbf{(c)} shows periodic density modulations and the relative phase profile \textbf{(d)} shows a staircase structure in space where the dotted lines mark $2\pi$ jumps representing the transitions between two adjacent solitons. The solitonic structure implies a density modulation, i.e  the the density ripples occur at the same positions of kinks in the phase difference profile degree of freedom.  The (in)homogeneous phase is called the (in)commensurate phase in the context of the Pokrovsky-Talapov transition. We refer to section ~\ref{sec:EFT} and the Appendix for more details on this notion.\label{fig:Commensurate} }
\end{figure*}

The minimum of the energy given in Eg.~\eqref{eq:lattice} is found by setting the
functional derivatives with respect to
$\alpha_1(x)$ and $\alpha_2(x)$ to zero
\begin{align}
 \frac{\delta {{E}}[{\alpha}]}{\delta {{\alpha}}^{\ast}_1({x})} &=0 \, , \quad  \frac{\delta {{E}}[{\alpha}]}{\delta {{\alpha}}^{\ast}_2({x})}=0 \, ,
\end{align}
which is explicitly given by
\begin{subequations}
\begin{align}
0 &= -  \partial^2_x \alpha_1 - \mu \alpha_1 + g |\alpha_1|^2 \alpha_1 -  e^{iQx}\alpha_2   \, , \label{eq:GPE1} \\
0 &= -  \partial^2_x \alpha_2 - \mu \alpha_2 + g |\alpha_2|^2 \alpha_2 -  e^{-iQx}\alpha_1 .   \label{eq:GPE2}
\end{align} \label{eq:GPE}
\end{subequations}
The equations \eqref{eq:GPE} together with the boundary conditions
\eqref{eq:BCMicro} form a boundary value
problem determining the field amplitudes ${\alpha}_b$.
In order to solve this boundary value problem we
study the imaginary time evolution
of the coherent state amplitudes given by
\begin{align}\label{imag}
\partial_\tau {\alpha}_b(\tau) = - \frac{\partial E}{\partial {\alpha}_b^{\ast}(\tau)}, \,
\end{align}
and its complex conjugate. For $\tau \rightarrow \infty$ we obtain a stationary value, i.e. $\partial_\tau {\alpha}_b(\tau) = 0$, which solves the system of differential equations  given by Eq.~\eqref{eq:GPE} and Eq.~\eqref{eq:BCMicro}.

Depending on the initial condition of \eqref{imag}, the imaginary time evolution may end
in a local minimum of the energy functional \eqref{eq:lattice}. The right choice of the initial condition will instead lead the
imaginary time evolution into the global minimum of equation \eqref{eq:lattice}.

We use the polar representation of the coherent amplitudes
\begin{align}\label{eq:pol}
{\alpha}_b(x) = \sqrt{n_b(x)} e^{i\theta_b(x)} \, ,
\end{align}
to obtain the density profile $n_b(x)$ and the phase difference
\begin{align}
    \theta_-(x) = \theta_1(x) - \theta_2(x)
\end{align}
and the \emph{shifted} relative phase
\begin{align}
\theta(x) = (2\pi)^{-1} \left[  \theta_-(x) - Qx \right]
\end{align}
determined by the stationary state of the imaginary time evolution. Atom-chip platforms can easily
access the phase difference by matter wave interference of the two Bose gases and absorption imaging of the density~\cite{Gring2012,NJP2013Smith}.

We observe that for all values of $Q$ the density drops to zero at the edges of the system as a result of the boundary conditions given in equation~\eqref{eq:BC}.
For small $Q$ the density profile is homogeneous in the center of the system whereas for larger $Q$
we observe  periodic density modulations  (see Fig.~\ref{fig:Commensurate}). 
Similarly, the \emph{shifted relative phase} $\theta(x)$ is homogeneous for small values of $Q$, but shows
a staircase structure for large values of $Q$.
The height of each jump is approximately $2\pi$; such jumps can be clearly seen  in the \emph{shifted} variable $\theta(x)$, rather than in the phase difference $\theta_-(x)$. We will call a single
phase jump of $2\pi$ a soliton and hence the profile $\theta(x)$ constitutes a staircase of solitons as can
be seen in Fig.~\ref{fig:Commensurate}d.

The qualitative difference of the density and
the \emph{shifted} phase profile for small and large $Q$ indicates a  phase transition from a homogeneous phase to an inhomogeneous phase.
In Sec.~\ref{sec:EFT} we will develop an effective description of the microscopic model in Eq.~\eqref{eq:HMicro}
and show that the phases can be effectively described by the Pokrovsky-Talapov model~\cite{PT, PT2, Bak}.

\subsection{Phase diagram}
\begin{figure}[t!]
\includegraphics[scale=.225]{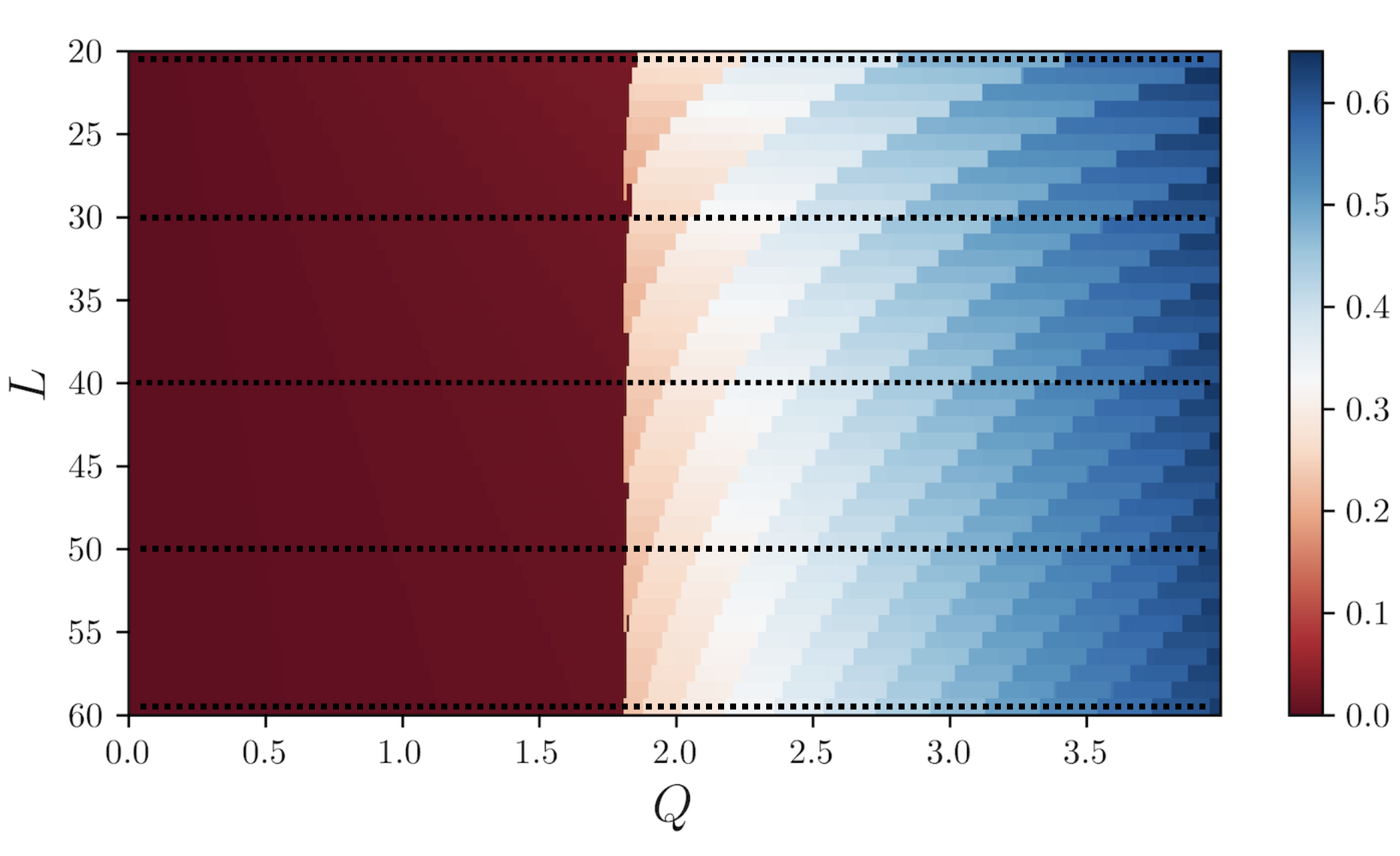}
\caption{\textbf{Density of solitons}: the homogeneous 
phase is characterized by a uniform phase profile and the absence of solitons (extended red region); around ${Q}\simeq 1.8$  solitons are injected with a non-vanishing density (incommensurate phase). The dotted lines mark different sizes of the system where  soliton injection is studied in Fig.~\ref{fig:Injection}. The microscopic parameters are the same as in Fig.~\ref{fig:Commensurate}.   \label{phdiagr}}
\end{figure}

In this section we discuss the order parameter, the phase 
diagram of the Hamiltonian~\eqref{eq:HMicro}. 
The number of
$(2\pi)$-jumps in the relative phase, the number of solitons,
\begin{align}
N_S = (2\pi)^{-1} \left[\theta(0) - \theta(L) \right]
\end{align}
is used to determine the density of solitons
\begin{align}
\rho=N_S/L \, ,
\end{align}
which acts as an order parameter of the transition.
The density of solitons as a function of $Q$ and $L$ is
depicted in Fig.~\ref{phdiagr}.

\begin{figure}[t!]
\includegraphics[scale=.56]{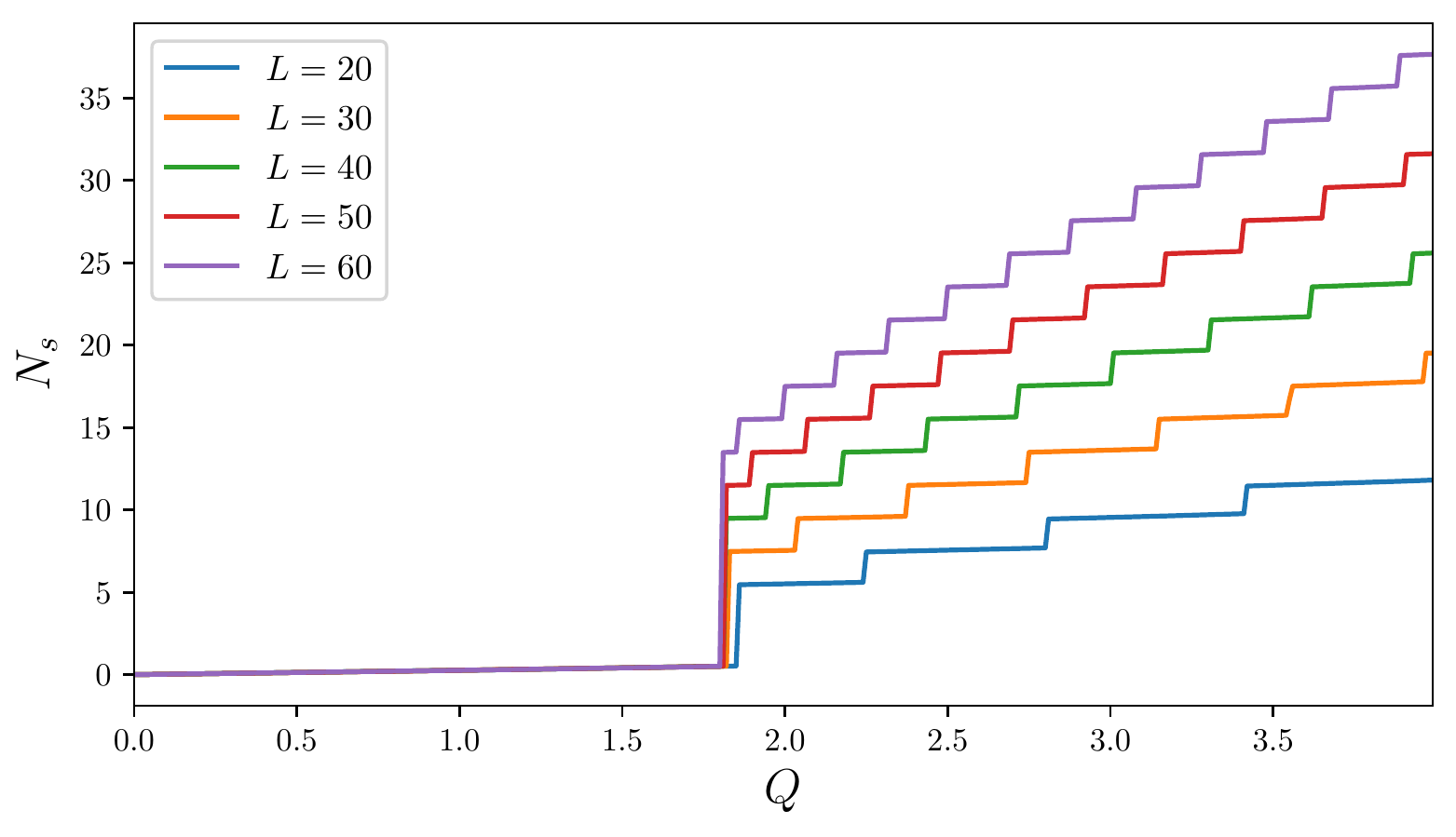}
\caption{ \textbf{Soliton injection}: For  finite system  size we observe discrete jumps in the number of solitons  upon increasing $Q$. This effect results from fitting  $m\simeq L/l_Q$  solitons  into the system with size $L$. The plots correspond to the horizontal cuts in the phase diagram of Fig.~\ref{phdiagr} (dotted lines). Microscopic parameters are the same as in Fig.~\ref{fig:Commensurate}.  }~\label{fig:Injection}
\end{figure}

The size of the system represents an additional length scale which  competes with the characteristic distance between two adjacent solitons $l_Q$. For an infinite system the distance between solitons diverges at the transition; on the other hand $l_Q$ is finite in the incommensurate phase, and a decreasing function of $Q$. The distance between two adjacent   solitons   scales as $l_Q\sim1/Q$. The competition between the system size $L$ and the soliton length $l_Q$ can introduce new features not present in the thermodynamic limit. For instance, Fig.~\ref{fig:Injection} shows the quantised injection of solitons as $Q$ is increased for various sizes of the system. 

Fig.~\ref{phdiagr} allows for for \emph{bona fide} estimate  of the critical point. 
For $L\simeq40$ and $Q\lesssim 1.8 = Q_c$, there is  negligible soliton density and identify this region as the homogeneous phase, whereas for $Q_c \gtrsim Q$ we obtain a non-vanishing soliton density, the inhomogeneous phase.  \\

\section{Bogoliubov theory}
\label{sec:bogo}

In this section we study  quantum fluctuations on top of the commensurate and incommensurate ground states. In particular we determine the Bogoliubov spectrum and discuss the physical features of the mode functions in the incommensurate phase.
We introduce small quantum fluctuations field operators  
on top of the mean field value
\begin{align}\label{exp}
\psi_b = \alpha_b + \delta \hat{\psi}_b \,.
\end{align}
Inserting this expansion into the Hamiltonian in Eq.~\eqref{eq:HMicro}
 and expanding  up to second order in  $\delta \hat{\psi}_b$, we obtain the Bogoliubov Hamiltonian 
\begin{widetext}
\begin{equation}\label{hambog}
H = \sum_{b=1}^2 \int_{0}^{L} dx |\partial_x \delta \hat{\psi}_b|^2
- \int_{0}^{L} dx \left( \delta \hat{\psi}^{\dagger}_1 e^{iQx} \delta \hat{\psi}_2 + h.c. \right) 
+  \frac{g}{2} \sum_{b=1}^2 \int_{0}^{L} dx   \left[ \alpha^2_b\delta \hat{\psi}^{\dagger}_b \delta \hat{\psi}^{\dagger}_b   +  (\alpha^{\ast}_b)^2 \delta \hat{\psi}_b \delta \hat{\psi}_b +4 |\alpha_b|^2  \delta \hat{\psi}^{\dagger}_b \delta \hat{\psi}_b   \right]
\end{equation}
Since the  field configuration in the ground state, $\alpha^{\ast}_b$, displays solitons, we have to solve an inhomogeneous Bogoliubov problem. The Heisenberg equations of motion for $\delta\hat{\psi}_b$ are
\begin{equation}\label{eq:mot}
\begin{split}
i \partial_{t} \delta \hat{\psi}_1 &= -\partial^2_x \delta \hat{\psi}_1
- e^{iQx} \delta \hat{\psi}_2
+ g (\alpha_1)^2  \delta \hat{\psi}^{\dagger}_1 +2 g |\alpha_1|^2 \delta \hat{\psi}_1,   \\
i \partial_{t} \delta \hat{\psi}_2 &= - \partial^2_x \delta \hat{\psi}_2
- e^{-iQx} \delta \hat{\psi}_1
+ g (\alpha_2)^2  \delta \hat{\psi}^{\dagger}_2 +2 g |\alpha_2|^2 \delta \hat{\psi}_2. 
\end{split}\end{equation}
These equations of motions can be solved by inserting
 the mode expansion
\begin{align}\label{expansion}
\delta \hat{\psi}_b(x,t) = \sum_m \left[ \hat{a}_m u_{m,b}(x) e^{-i\omega_{m} t} + \hat{a}^{\dagger}_m v^{\ast}_{m,b}(x) e^{i\omega_{m}t} \right] \,
\end{align}
 in the equations of motion~\eqref{eq:mot}; this procedure yields  the eigenvalue problem
\begin{equation}\label{bogol}
\omega_m
\begin{pmatrix}
u_{1,m} \\
u_{2,m} \\
v_{1,m} \\
v_{2,m}
\end{pmatrix}
=
\begin{pmatrix}
h_1(x)                 & -e^{iQx}                & g(\alpha_1)^2 & 0 \\
-e^{-iQx}              & h_2(x)                  & 0             & g (\alpha_2)^2 \\
-g (\alpha^{\ast}_1)^2 & 0                       & -h_1(x)       & e^{-iQx} \\
0                      & -g(\alpha^{\ast}_2)^2   & e^{iQx}       & -h_2(x)
\end{pmatrix}
\begin{pmatrix}
u_{1,m} \\
u_{2,m} \\
v_{1,m} \\
v_{2,m}
\end{pmatrix} \, ,
\notag
\end{equation}
where $h_b(x) = -\partial_x^2 + 2g |\alpha_{b}|^2$.
The Bogoliubov modes satisfy 
\begin{align}
    \delta_{nm} = \sum_x \left[ u^{\ast}_n(x) u_m(x) - v^{\ast}_n(x) v_m(x) \right] \, ,
\end{align}
which ensures the canonical commutation relation of $\hat{a}_m$ and $\hat{a}^{\dagger}_m$.

\begin{figure}[t!]
\includegraphics[scale=0.45]{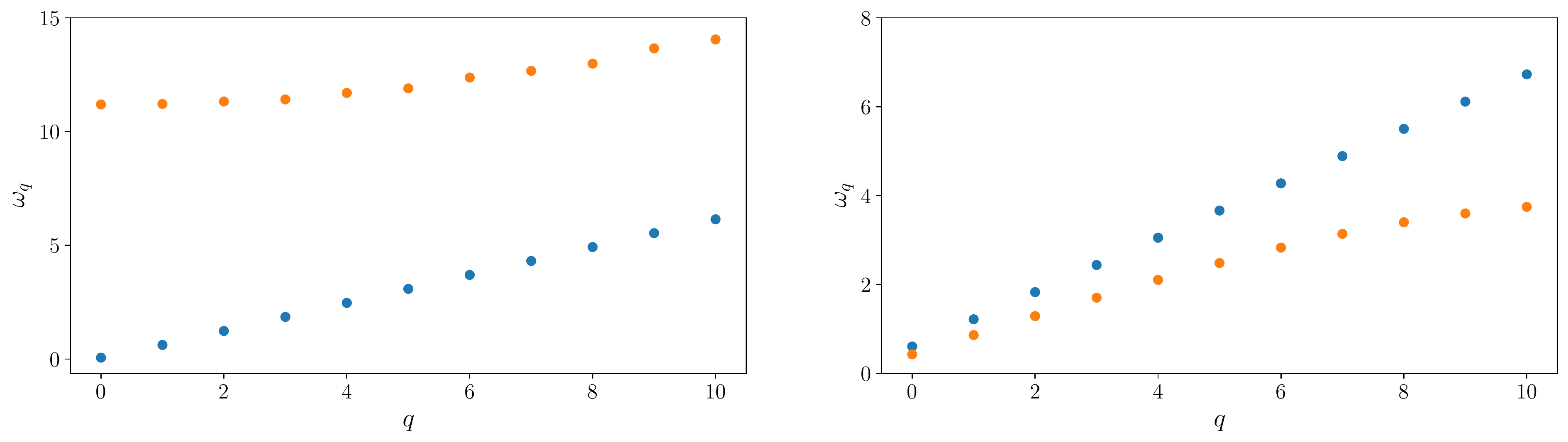}
\caption{ \textbf{Dispersion Relation}  Left panel: dispersion relation of the 'symmetric' (orange) and 'anti-symmetric' (blue) modes in the commensurate phase ($Q=0.9$). The dispersion relation of the  'anti-symmetric'  modes as a function of momentum is quadratic at low energies with a gap $\Delta= \sqrt{2J(2J+2\mu)}\simeq 11$ (here $J=1$ and $\mu=30$). Right panel: In the incommensurate phase ($Q=2.1$) we observe  two linearly dispersing sound  modes at low energies. These two sound modes  correspond to the $U(1)$ phonon  (blue) and  to the 
soliton phonon (orange). A scaling with system size of the low energy eigenvalues is provided in the Appendix.
\label{figDisp}}
\end{figure}

\end{widetext}

\begin{figure*}[t!]
\includegraphics[width=\textwidth]{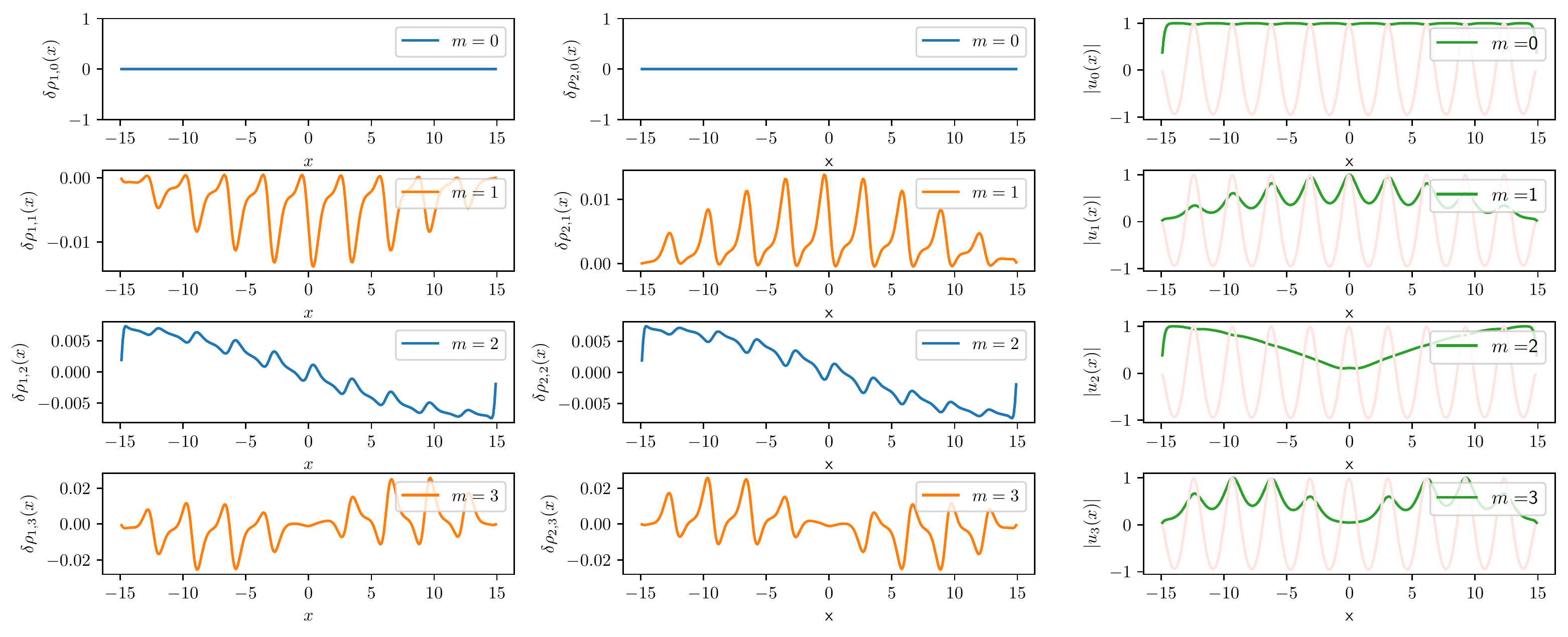}
\caption{
Columns (1) and (2): spatial profile of the density fluctuations determined from the Bogoliubov calculation for each mode. 
The two columns allow to categorize the modes 
within tube 1 and 2 into symmetric and anti-symmetric modes.
Symmetric modes are labeled in blue, while anti-symmetric modes are labeled in orange. For instance, the third and fourth rows display the first excited state in the symmetric and anti-symmetric sectors. By  sorting the eigenvalues according to this symmetry considerations we show the dispersion relation of the symmetric (blue) and anti-symmetric (orange) modes  in the right panel of Fig.~\ref{figDisp}. Column (3) shows the relation between the ground state current (light red) and 
the absolute value of the Bogoliubov mode 
functions (green). Note that the relative phase and the ground state current are peaked everywhere a soliton is formed in the relative phase profile of the two tubes; the modes in the anti-symmetric sector are sensitive to the presence of solitons and are peaked at the same positions as the solitons. In the figure we have plotted the absolute value of the Bogoliubov mode functions (green) to illustrate the influence
of the modulation of the background field on the mode-functions. The modes of the symmetric sectors show a milder spatial corrugation (see $m=2$). This is in contrast with the   Pokrovsky-Talapov field theory where the symmetric modes are completely insensitive to soliton injection, because they do not couple to the anti-symmetric modes.
\label{fig6}}
\end{figure*}

We numerically solve the system in equation~\eqref{bogol} and
obtain the spectrum and mode functions. For $Q=0$ the Hamiltonian of the tunnel-coupled Bose gases is symmetric under the exchange of the tube labels $1\leftrightarrow 2$; this allows to diagonalize the system in two independent subspaces
corresponding to  the two parities  associated to the symmetry. In the following we will call  symmetric and anti-symmetric the modes
belonging respectively to these two subspaces. The dispersion relation of the  anti-symmetric degrees of freedom is particle-like: it is known to have a gap  $\Delta= \sqrt{2J(2J+2\mu)}$, and to grow parabolically at low momenta~\cite{PhysRevA.68.053609}. 
When the background density of the condensates is flat, the tunnelling operator gaps only the anti-symmetric modes~\cite{Ima, imambekov2007fundamental}, while the 
 spectrum of the symmetric degrees of freedom remains gapless at low energies,
following the linear dispersion relation of the conventional  Bogoliubov theory for homogeneous gases~\cite{PhysRevA.68.053609}. %
We find by the numerical solution of~\eqref{bogol} that similar results hold for a pair of tunnel coupled Bose gases if $Q<Q_c$ (homogeneous phase). In particular, one can 'adiabatically' connect the $Q=0$ and the $Q<Q_c$ state in the numerical evaluation of the eigenvalues of equation~\eqref{bogol}.
Hence  the spectrum at $Q<Q_c$ can still be  separated into two  distinct branches with  properties analogue  to the $Q=0$ case. In particular the branches persist despite the  '$1\leftrightarrow 2$' symmetry is explicitly  broken by a non-vanishing value of $Q$; this  is illustrated in left panel of  Fig.~\ref{figDisp}. 

At $Q=Q_c$ the system experiences a commensurate-incommensurate phase transition of first order which closes the gapped mode of the anti-symmetric sector~\cite{arist, Bak} (when fluctuations  are included, the transition is expected to become of second order, however, our mean-field ansatz for the ground state, and the associated phase diagram~\ref{phdiagr}, does not include such fluctuations~\cite{PT, PT2}).
Accordingly  we expect two linear sound
modes in the inhomogeneous phase. One branch resulting from  the breaking of $U(1)$ symmetry, while 
the second branch is due to the breaking of continuous translation invariance into a discrete translation symmetry, as a result of the formation of a soliton lattice spacing  $l_Q$ in the system. Therefore, these two phonon branches belong respectively  to the symmetric and anti-symmetric sectors of the inhomogeneous Bogolyubov problem. The speed of sound of the soliton phonon is known to follow linearly the misfit parameter $Q$~\cite{arist, Bak}.

In the realistic quantum simulator of the PT field theory studied in this work  the  symmetric and anti-symmetric 
branches  of the dispersion relation hybridize for $Q>Q_c$, since the  profile of the  density of the two gases displays spatial modulations, contrary to the  the theory of tunnel-coupled Bose gases developed in~\cite{Ima}). This  coupling effect is moderate in our model as can be inferred from the depth of the density ripples in Fig.~\ref{fig:Commensurate}, but it is completely absent in the conventional PT theory which can be  written as a sole function of  the  phase difference of the two gases~\cite{arist}. Since the latter constitutes an anti-symmetric degree of freedom,  the soliton phonon is the only relevant soft mode in  PT field theory. 
In Fig.~\ref{figDisp} we plot the low energy eigenvalues and find evidence for the two sound modes. 
For low-lying energy eigenvalues we obtain the separation 
into symmetric and anti-symmetric branches by studying the spatial structure of small density fluctuations
\begin{equation}\label{modes}
\delta\rho_{b}(x) =  \alpha_b(x) \delta \hat{\psi}^\dag_{b}(x)+\text{h.c.} \, ,
\end{equation} 
which allows to split the  mode functions  into two groups  
distinguished by their dominant waveform (see left panel of Fig.~\ref{fig6}). In terms of the operator~\eqref{expansion} we can rewrite Eq.~\eqref{modes} as
 
\begin{align}
	\delta\rho_{b,m}(x) \equiv 	\alpha_b(x) \left[\alpha_m u_{m,b}(x)  + \alpha^{\dagger}_m v^{\ast}_{m,b}(x) \right] +\text{h.c.}
\end{align}
in terms of the coherent state amplitude $\alpha_b(x)$ in the ground state of the tube $b$. 

We observe that the excitations in the anti-symmetric sector preserve memory of the location of the solitons forming in the ground state. In order to illustrate that, we compare the ground state current 
\begin{equation}
j_b(x)\equiv i \left[ \alpha^{\ast}_b(x) \partial_x \alpha_b(x)-\text{h.c.} \right]
\end{equation}
with the  profile of the mode functions in the right panel of Fig.~\ref{fig6}. From Fig.~\ref{fig:Commensurate} one can infer that whenever a soliton appears as a density ripple/phase jump, the current has a peak in a one-to-one correspondence with a peak in the mode functions' profiles. This connection explains the strong spatial variations displayed by the anti-symmetric modes when compared to  symmetric ones: solitons pin quantum fluctuations which in turn model the shape of the mode-functions. Conversely, the mode functions of the symmetric phase show a more regular spatial profile since they are weakly coupled to solitons.

\begin{figure}[t!]
\includegraphics[scale=.6]{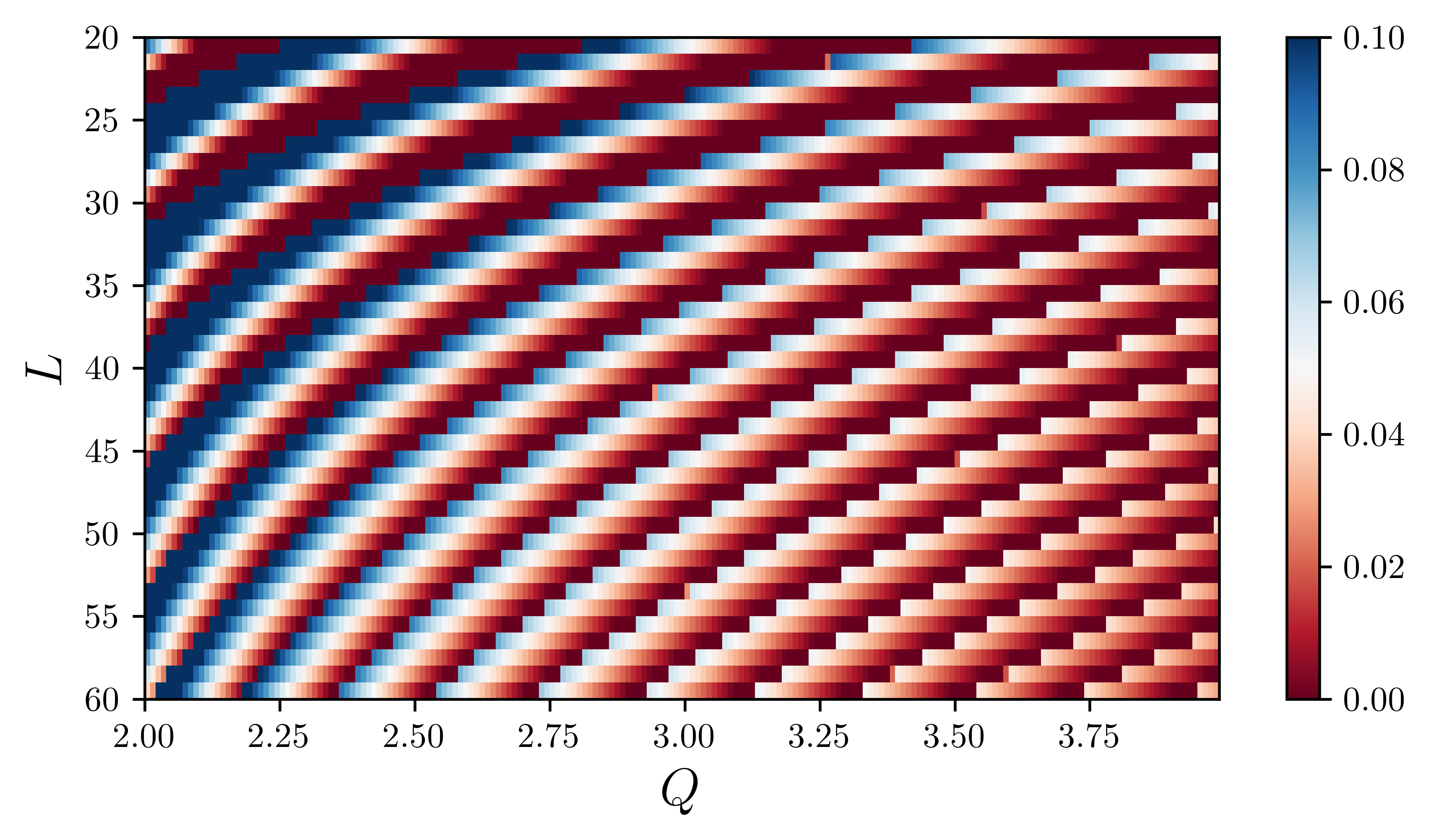}
\caption{ \textbf{Comparison with the  Pokrovsky-Talapov model}: Plot of ${|\rho_{\textit{mic.}}-\rho_{\textit{PT}}|}/{\rho_{\textit{mic.}}}$, with $\rho_{\textit{mic.}}$  the soliton density of the microscopic model studied in this work, and $\rho_{\textit{PT}}$  the soliton density of the effective PT field theory. A mismatch between $\rho_{\textit{PT}}$ and $\rho_{\textit{mic.}}$  indicates regions of the phase diagram, where corrections beyond the conventional PT description are present (microscopic parameters here are the same as in Fig.~\ref{fig:Commensurate}).  Such corrections are more pronounced close to the critical point and vanish as the system size increases. \label{fig:PD2} }
\end{figure}

\section{Effective field theory \label{sec:EFT}}
In this section we derive an effective field theory for the Hamiltonian
given in Eq. \eqref{eq:HMicro} following the lines of Ref.~\cite{gritsev}. Therefore we express the coherent amplitudes
${\alpha}_b(x)$ by density $n_b(x)$ and phase $\theta_{b}(x)$, and
 obtain the energy
\begin{align}
E =&  \sum_{b=1}^2 \int_0^L dx \left[ \frac{(\partial_x n_b)^2}{4n_b} + n_b (\partial_x \theta_b)^2  - \mu n_b  +  \frac{g}{2} n^2_b   \right] \notag \\
& - 2 \int_0^L dx \sqrt{n_1 n_2} \cos \left[ \theta_1 - \theta_2 -Qx \right]    \, .
\label{eq:Energy2}
\end{align}
The boundary conditions \eqref{eq:BCMicro} become
\begin{align}\label{microbc}
n_b(x=0) = n_b(x=L) = 0 
\end{align}
with a free phase angle at the boundary.
Expanding the density in Eq.~\eqref{eq:Energy2}  as
\begin{align}\label{eq:smallden}
n_b(x) = n_0 + \delta n_b(x),
\end{align}
around a homogeneous background density $n_0$, we obtain
\begin{align}\label{eq:zeroord}
E =  n_0 \int_0^L dx  \left[ \sum_{b=1}^2 (\partial_x \theta_b)^2 - 2 \cos \left( \theta_1 - \theta_2 -Qx \right) \right]   \, .
\end{align}
The expansion around a homogeneous background density is justified as can be seen
from Fig. \ref{fig:Commensurate} which shows that fluctuations  $\delta n_b(x)$ are small besides at the edges of the system.
The central and relative coordinates
\begin{subequations}
\begin{align}
\theta_+ &= \tfrac{1}{2} (\theta_1 + \theta_2), \\
\theta_-   &=  \theta_2 - \theta_1,
\end{align}
\end{subequations}
decouple, and the energy density can be written as
\begin{align}~\label{eg:modelPT1}
E = n_0 \int_0^L dx \left[\frac{1}{2} (\partial_x \theta_-)^2 -2 \cos(\theta_- - Qx) \right] \, .
\end{align}
Changing variables to $\theta = \theta_- - Qx$, we find
\begin{align}~\label{eq:modelPT2}
E = n_0  \int_0^L dx \left[\frac{1}{2} (\partial_x \theta)^2 + Q\partial_x \theta -2 \cos\theta \right] \, ,\end{align}
 which corresponds to the energy density of the classical Pokrovsky-Talapov model (see Appendix). 
The Pokrovsky-Talapov model~\eqref{eq:modelPT2} describes the transition from a commensurate phase with a homogeneous ground state to an incommensurate phase characterized by a finite soliton density. 

At the mean field level and for a large system size the critical point is located at $\bar{Q}_c= 4\sqrt{2}/\pi\sim1.8$. In an atom-chip experiment with $^{87}$Rb and a typical effective Raman assisted tunneling strength $|\widetilde{J}|\sim 2\pi \times 1$Hz  this results in a critical point located at $ \bar{Q}_c/ x_0 \approx 0.1 (\mu$m$)^{-1}$.
Contrary to the Pokrovsky-Talapov model the relative phase degree of freedom in our model couples to  variations of the density $\delta n_b(x)$  contributing with terms beyond the leading order expansion  in Eq.~\eqref{eq:zeroord}. These terms are  responsible for effects beyond the Pokrovsky-Talapov effective description. 

It is therefore natural to investigate to which extent the microscopic model reproduces Pokrovsky-Talapov physics.
 The agreement of the order parameter between the microscopic model  and the effective field theory is excellent in the commensurate phase, and  far away from the phase transition in the incommensurate phase, as  illustrated in  Fig.~\ref{fig:PD2}. 

\section{Conclusions}

In this work we have presented  an ultra-cold atom system which can be employed as a simulator of a commensurate/incommensurate phase transition. Specifically we proposed to use Raman tunnel-coupled, one-dimensional quantum liquids in atom chip experiments as a platform to study the quantum effects in the incommensurate phase. We have shown how this model can be understood as an approximate Pokrovsky-Talapov model, and included quantum fluctuations within an inhomogeneous Bogoliubov calculation.
We have investigated the differences between the PT field theory and the realistic platform considered in our work, discussed deviations from the PT commensurate/incommensurate phase diagram, and studied how quantum features interplay with solitons and influence the shape of the wave-functions. The control offered by this platform allows to test regimes of validity of the effective field theory, and it paves the way to a number of future directions.   A non-equilibrium study of the Pokrovsky-Talapov model on a atom chip could offer interesting perspectives, since such platforms have been already shown to represent formidable simulators for the dynamics  of tunnel coupled Luttinger liquids~\cite{Gring2012, Langen2013, Gring2015, Science2018Rauer}.
~Future work could encompass the dynamical production and annihilation of solitons by quenching the misfit parameter across the phase transition, or by studing  light-cone propagation of correlation functions in presence of multiple speeds of sounds.  This could motivate  a novel series of experiments involving non-equilibrium dynamics of topological excitations, as kinks, in one-dimensional quasi-condensates. We see  our results as a intermediate step towards a surge of novel interest on quantum simulations in and out-of-equilibrium of the Pokrovsky-Talapov physics.
Our study can be also straightforwardly generalized to a large class of quantum spin chains~\cite{PhysRevB.68.134431, PhysRevLett.68.3889, PhysRevLett.81.910}, to two component Bose mixtures with spin, and to the XXZ spin chain with magnetic field, which maps into the PT model~\cite{giamarchi2003quantum}. 

\section{Acknowledgements}
We thank discussions with R. Citro, M. Knap, M. Lewenstein, R. Schmidt, D. Sels, L. Tarruell.
VK was supported by a Feodor Lynen Fellowship of the Alexander von Humboldt Foundation and received funding from the European Union's Horizon 2020 research and innovation programme under the Marie Sk³odowska-Curie grant agreement No. 754510           (PROBIST).
JM was supported by the European Union's Framework Programme for Research and Innovation Horizon 2020 (2014-2020) under the Marie Sklodowska-Curie Grant Agreement No. 745608~('QUAKE4PRELIMAT').
{SJ is supported by the Erwin Schroedinger Quantum Science \& Technology (ESQ) Fellowship, which has received funding from the European Union's Horizon 2020 research and innovation programme under the Marie Sklodowska-Curie grant agreement No. 801110~('1D-AGF').}
This work is supported by the Harvard-MIT CUA,
the AFOSR-MURI Photonic Quantum Matter (award FA95501610323),
the DARPA DRINQS program (award D18AC00014), the Austrian Science Fund (FWF) through the Austrian participation to the German SFB 1225 'ISOQUANT'.

\clearpage

\begin{widetext}

\section*{Appendix: The Pokrovsky-Talapov model}

The Hamiltonian density of the classical, one dimensional, Pokrovsky-Talapov model is a sine-Gordon  model
with a spatially dependent cosine term
\begin{align}
    E =  \int_0^L dx \left[\frac{1}{2} (\partial_x \theta_-)^2 -2 \cos(\theta_- - Qx) \right] \, . \label{eq:PTModel1}
\end{align}
Frequently, one performs a change of variables
\begin{align}
    \Theta = \theta_- - Qx \, ,
\end{align}
such that the energy becomes
\begin{align}
E = \int_0^L dx \left[\frac{1}{2} (\partial_x \Theta)^2 + Q \partial_x \Theta  -2 \cos \Theta  \right] \, ,
\label{fieldtheory}\end{align}
where the spatial gradient term proportional to $Q$ fixes the density of solitons~\cite{arist} and we left
out a constant term.
The Pokrovsky-Talapov model hosts a commensurate-incommensurate transition characterised by the onset of a
non-vanishing density of solitons occurring for $Q>Q_c$ where $Q_c$ is the critical mismatch parameter,
whereas for $Q<Q_c$ the field configuration is homogeneous.
A qualitative  way to understand the commensurate-incommensurate transition, is realizing that for large $J$ (cf. Eq.~\eqref{en}; in Eq.~\eqref{eq:PTModel1} we have set $J=1$), the potential tends to favour minima of the potential, $\Theta_n=2n\pi$, with $n$ an integer;  on the other hand for large $Q$ the gradient term becomes the dominant contribution to the energy and a field configuration following the linear trend $\Theta\simeq Qx$ is favoured, on top of which a soliton staircase structure  establishes. The competition between these two energetically different configurations leads to the onset of the commensurate-incommensurate transition at $Q=Q_c$.

In order to determine the ground-state field configuration we take the derivative of \eqref{eg:modelPT1}
\begin{align}
    0 = \frac{\delta E }{\delta \theta_-(x)} \, ,
\end{align}
which leads to
\begin{align}
    0 = -\partial^2_x \theta_- + 2 \sin (\theta_- -Qx ) \,.
\end{align}
Changing again the variables to $\Theta = \theta_- - Qx$ we obtain the equation
\begin{align}
    0 = \partial^2_x \Theta + 2 \sin  \Theta  \, , \label{eq:EOM_PT}
\end{align}
where we have shifted $\sin\Theta\to\sin(\Theta+\pi)$ in~\eqref{eq:PTModel1}, as also done in ~\cite{arist}.\\

Multiplying from the right hand side with $\partial_x \Theta$ we obtain
\begin{align}
    0 = \partial_x \left[ \frac{1}{2}  (\partial_x \Theta)^2 +  2 \cos \Theta \right] \, .
\end{align}
Integrating the last equation, this yields
\begin{align}
    \frac{C}{2} = \frac{1}{2}  (\partial_x \Theta)^2 -  2 \cos \Theta
\end{align}
with an integration constant $C$. Integrating once again results in
\begin{align}
    x - x_0 = \int_{\Theta(x_0)}^{\Theta(x)} \frac{d \Theta}{\sqrt{C + 4\cos \Theta } } \, .
\end{align}
Since we consider an infinitely long system we can set without loss of generality $\theta(0) = 0$ and obtain
\begin{align}
    x  = \int_{0}^{\Theta(x)} \frac{d \Theta}{\sqrt{C + 4\cos \Theta } } \, . \label{eq:Soliton0}
\end{align}
Solving this equation for $\Theta(x)$ leads to
\begin{align}
    \Theta(x) = 2 \textsl{Am} \left( \sqrt{C/4+1} ~x, \frac{2}{1+C/4} \right),
\end{align}
where the right hand side is determined by inverse Jacobi amplitude, $Am(z,k)$, of argument $x$ and index $k$.

\subsection{$l_Q$ and location of the critical point}

We  insert the relation~\eqref{eq:Soliton0} into the energy of the PT model and  consider the phase offset ($\Delta\Theta=2\pi$) given by the injection of a single soliton in the system;  we obtain
\begin{align}\label{energy}
    E/L = -{2\pi Q}+ \int_0^{2\pi} d\Theta \sqrt{C+4\cos\Theta}=-2\pi Q +2\left[\sqrt{C-4}~\mathcal{E}\left(\frac{8}{4-C}\right)+\sqrt{C+4}~\mathcal{E}\left(\frac{8}{C+4}\right)\right] \, ,
\end{align}
where $\mathcal{E}(\cdot)$ is the  complete elliptic integral of the second kind, and the integration constant $C$ has yet to be determined.
\\

The location of the critical point, $Q_c$, can be  calculated determining when the 'chemical potential' $Q$ makes the  solitonic configuration energetically favourable~\cite{laz}: normally the energy of a soliton is higher than the minimal energy configuration, $E_n=0$, of a sine-Gordon field theory, corresponding to $\Theta_n=2n\pi$ (field pinned at the minima of the cosine). This excess of energy can be compensated by the Pokrovsky-Talapov misfit, $ Q$; analogously to a chemical potential, it can lower the energy of a soliton, which can  become  a favourable energy configuration when its energy equals that one of the field in the commensurate phase. Since we set the latter equal to zero, this corresponds at the vanishing of the expression in~\eqref{energy}:

\begin{align}\label{criticest}
2\pi Q =2\left[\sqrt{C-4}~\mathcal{E}\left(\frac{8}{4-C}\right)+\sqrt{C+4}~\mathcal{E}\left(\frac{8}{C+4}\right)\right]
\end{align}

 The length, $l_Q$, diverges at $C=4$, and it is defined as
\begin{align}
    l(C)  = \int_{0}^{2\pi} \frac{d \Theta}{\sqrt{C + 4\cos \Theta } } \, . \label{eq:length0}
\end{align}
 Intuitively, this is a signature of the onset of the commensurate-incommensurate transition, since, upon increasing $Q$, the density of solitons increases and therefore their mean spacing  will decrease. This allows already to determine the location of the critical value of $Q$; replacing $C=4$ into~\eqref{criticest} we find $Q_c=4\sqrt{2}/\pi$.\\

Alternatively, using the  symmetry of the integrand we get
\begin{align}
    l(C) = 2 \int_0^{\pi} \frac{d \Theta}{\sqrt{C + 4\cos \Theta } } \label{eq:length1}.
\end{align}
Setting $C = 4 + \varepsilon$, with $\varepsilon\ll1$, we obtain

\begin{align}\label{eq:length2}
    l(\varepsilon) = - \frac{\sqrt{2}}{4} \log \varepsilon + ...
\end{align}

On the other hand we have from Eq.~\eqref{criticest}
\begin{align}
    Q = \frac{1}{2\pi} \int_0^{2\pi} du \sqrt{4 + \varepsilon +4\cos u} \label{eq:Q_eps}
\end{align}
Close to the critical point, $C=4+\varepsilon$ (with $\varepsilon \ll 1$),
 we obtain
\begin{align}
    Q = \frac{1}{2\pi} \int_0^{2\pi} \sqrt{4 + 4\cos u} + \frac{1}{4\pi} \int_0^{2\pi} \frac{du}{\sqrt{4+\varepsilon+4\cos u}}\varepsilon + \mathcal{O}(\varepsilon^2)
\end{align}
The first integral defines $Q_c$ and the second integral yields
\begin{align}\label{critic}
    Q = Q_c + \frac{l(\varepsilon)}{4\pi}  \varepsilon, \,
\end{align}
with $Q_c=4\sqrt{2}/\pi$.
Solving \eqref{eq:length2} for $\varepsilon\ll1$,   leads to
\begin{align}
    \varepsilon = e^{-\frac{4}{\sqrt {2}}l(\varepsilon)} \, ,
\end{align}
and inserting this result into Eq.~\eqref{critic}, we find
\begin{align}
    Q - Q_c =
 \frac{1}{4\pi}l(\varepsilon) e^{-\frac{4}{\sqrt {2}}l(\varepsilon)}.
\end{align}
Solving for $l(\varepsilon)$ close to criticality ($\varepsilon\ll1$) we obtain~\cite{solitbook}
\begin{align}
    l(\varepsilon) \propto {\log \frac{1}{Q-Q_c}}.
\end{align}
This result implies that the density of the 'kink condensate' ${\rho}$ (the inverse of the solitons' spacing) vanishes logarithmically close to the transition~\cite{arist} with diverging derivative
\begin{equation}\label{eq:dens}
{\rho}\propto\left(\log\frac{1}{Q-Q_c}\right)^{-1}, \quad Q\simeq Q_c;
\end{equation}
  it effectively grows linearly, ${\rho}\propto Q$, for  $Q > 2 Q_c$, as reported in Ref.~\cite{arist}. The density of solitons determine the size of the steps in the solitonic staircase shown in Fig.~\ref{figplat}, $l_Q\propto 1/{\rho} \propto \log(1/(Q-Q_c))$.
Over intervals of size $l_Q$, the function $\Theta(x)$ assumes practically constant value $2n\pi$ and then over an interval of size $l_K$ it jumps by $2\pi$.
 In the regime where the density grows linearly with the chemical potential, the characteristic size of a soliton becomes then $l_Q\sim1/Q$.

 \subsection*{Appendix: Finite size scaling of the lowest energy Bogoliubov eigenvalues}

The right panel of Fig.~\ref{spectrum} shows the $1/L$ linear scaling of the  low energy eigenvalues in 
the symmetric sector of the homogeneous phase. 

In the left panel of Fig.~\ref{spectrum} we plot the finite size scaling of the low energy eigenvalues in the incommensurate phase. We  observe $1/L$ scaling for  certain ranges of  $L$ interrupted by 'jumps'. Such discontinuities are understood as new solitons'  injections  when
 $L$ is increased (cf. with  Figs.~\ref{phdiagr} and~\ref{fig:Injection}). This injection of solitons leads to a 
readjustment of the  background field in the incommensurate phase and therefore to a shift in the slope of the $1/L$ fit of the low energy eigenvalues. This represents 
an other imprint of the solitons on the quantum properties of our model.

\begin{figure}[h!]
\includegraphics[width = \textwidth]{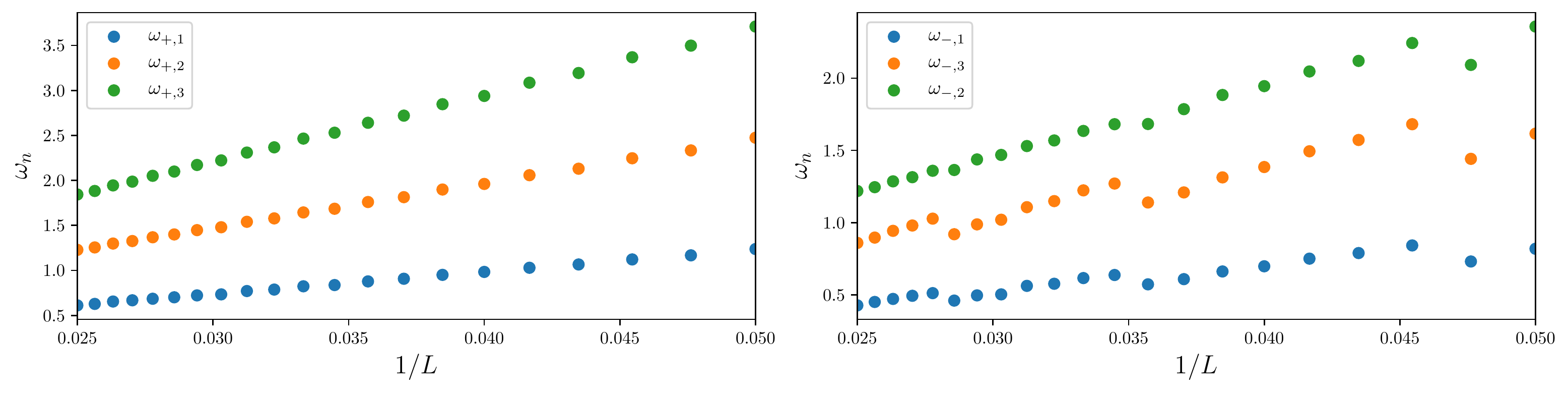}
\caption{\textbf{Finite $1/L$ scaling} of the lowest eigenvalues of the Bogoliubov spectrum in the inhomogeneous phase  (left panel, $Q=2.1$) and in the symmetric sector of the homogeneous phase (right panel, $Q=0.9$). The lowest energy eigenvalue crosses the horizontal axis around $L\simeq60$ in both phases.
\label{spectrum}}
\end{figure}

\hphantom{bbbbbbbbbbbbbbbbbbbbbbbbbbbbbbbbbbbbbbbbbbbbbbbbbbbbbbbbbbbbbbbbbbbbbbbbbbbbbbbbbbbbbbbbbbbbbbbbbbbbb
 bbbbbbbbbbbbbbbbbbbbbbbbbbbbbb}\hphantom{bbbbbbbbbbbbbbbbbbbbbbbbbbbbbbbbbbbbbbbbbbbbbbbbbbbbbbbbbbbbbbbbbbbbbbbbbbbbbbbbbbbbbbbbbbbbbbbbbbbbb
 bbbbbbbbbbbbbbbbbbbbbbbbbbbbbb}
 \hphantom{bbbbbbbbbbbbbbbbbbbbbbbbbbbbbbbbbbbbbbbbbbbbbbbbbbbbbbbbbbbbbbbbbbbbbbbbbbbbbbbbbbbbbbbbbbbbbbbbbbbbb
 bbbbbbbbbbbbbbbbbbbbbbbbbbbbbb}
\end{widetext}
\newpage
\bibliography{biblio_PT}

\end{document}